\documentclass[a4paper]{article}
\usepackage[T1]{fontenc}
\usepackage[utf8]{inputenc}
\usepackage[english]{babel}

\usepackage{enumitem}
\usepackage{graphicx}
\usepackage{color}
\usepackage[dvipsnames,svgnames*,x11names*]{xcolor}
\usepackage[normalem]{ulem}
\usepackage{subfig}
\usepackage{url}
\usepackage{booktabs}
\usepackage{arydshln}
\usepackage{multirow}
\usepackage{siunitx}
\usepackage{geometry}
\usepackage[labelfont={sf,bf}]{caption}
\usepackage{setspace}
\usepackage{amsmath, amssymb}
\usepackage{bm}

\usepackage{xpatch}
\usepackage[autostyle,italian=guillemets]{csquotes}
\usepackage[backend=biber,style=authoryear-comp, sortcites, sorting=ynt, maxcitenames=1, uniquelist=false, backref=true, maxbibnames=99]{biblatex}
\usepackage{guit}
\DeclareDelimFormat{nameyeardelim}{\addcomma\space}

\usepackage{xpatch}
\usepackage{xurl}
\usepackage{hyperref}
\hypersetup{
  pdftitle={Measurement Error Models for Spatial Network Lattice Data: Analysis of Car Crashes in Leeds},
  pdfauthor={Andrea Gilardi, Riccardo Borgoni, Luca Presicce and Jorge Mateu},
  colorlinks=true,
  linkcolor=RoyalPurple,
  filecolor=Maroon,
  citecolor=Black,
  urlcolor=Blue
 }

\makeatletter
\def\adl@drawiv#1#2#3{%
        \hskip.5\tabcolsep
        \xleaders#3{#2.5\@tempdimb #1{1}#2.5\@tempdimb}%
                #2\z@ plus1fil minus1fil\relax
        \hskip.5\tabcolsep}
\newcommand{\cdashlinelr}[1]{%
  \noalign{\vskip\aboverulesep
           \global\let\@dashdrawstore\adl@draw
           \global\let\adl@draw\adl@drawiv}
  \cdashline{#1}
  \noalign{\global\let\adl@draw\@dashdrawstore
           \vskip\belowrulesep}}
\makeatother

\title{Measurement Error Models for Spatial Network Lattice Data: Analysis of Car Crashes in Leeds}
\author{Andrea Gilardi, Riccardo Borgoni, Luca Presicce and Jorge Mateu}
\date{}

\begin{document}

\newrefcontext[sorting=ynt]

\maketitle
\setstretch{1.25}

\begin{abstract}
Road casualties represent an alarming concern for modern societies. During the last years, several authors proposed sophisticated approaches to help authorities implement new policies. These models were usually developed considering a set of socioeconomic variables and ignoring the measurement error, which can bias the statistical inference. This paper presents a Bayesian model to analyse car crashes occurrences at the network-lattice level, taking into account measurement error in the spatial covariate. The suggested methodology is exemplified by considering the collisions in the road network of Leeds (UK) during 2011-2019.  Traffic volumes are approximated using an extensive set of counts obtained from mobile devices and the estimates are adjusted using a spatial measurement error correction. 
\end{abstract}

\section{Introduction}
\label{sec:intro}

According to the World Health Organisation \parencite{WHO2018, WHO2020}, the global number of road casualties is steadily increasing since 2016 and, in the last few years, reached unacceptably high levels. Car crashes are the leading cause of death for children and young people aged 5-29 years and the eighth cause of death in all age groups. Traffic injuries have direct social costs and indirect economical consequences (such as an adverse impact on the burden of hospitalisation or an increased health expenditure) that represent, on average, 3\% of the annual GDP \parencite{WHO2021}. These problems are slightly less severe for more developed countries (mainly Europe or North America), but the situation is still alarming considering that people from lower socio-economic backgrounds are more likely to be involved in traffic casualties. Moreover, the burden of road accidents is disproportionately carried by vulnerable road users (such as pedestrians or cyclists) and, according to a mental health study, $39.2$ per cent of car crash survivors develop post-traumatic stress disorder and require psychological assistance \parencite{blanchard1995psychiatric}. Therefore, local and global authorities define road safety plans \textit{``an unfinished agenda"}, demanding innovative approaches and evidence-based interventions \parencite{WHO2021}.

For these reasons, starting from the last century, several academic researchers developed systematic approaches to analyse traffic injuries and help decision-makers implement new policies to mitigate the problem. The first papers focused on a descriptive analysis of the crashes frequencies \parencite{smeed1949some}, whilst, during the last two decades of the last century, more advanced statistical techniques, such as linear and generalised linear models, were proposed \parencite{miaou1993modeling, miaou1994relationship}. Since the beginning of the 2000s, the statistical models have started taking into account the geographical nature of the road collisions, improving the estimation process thanks to the spatial autocorrelation in the observed variables \parencite{el2009urban, loo2015spatial}. At the time of writing, the two most popular approaches in road safety analysis involve aggregation of point-level car crash data to larger spatial units, either polygonal areas (such as census blocks or traffic analysis zones) or linear network features (such as street segments or corridors and intersections). In both cases, the spatial support can be described as a lattice, possibly supplemented by a neighbourhood matrix.

The first papers on spatial road safety analysis were developed using the areal approach \parencite{miaou2003roadway, aguero2006spatial, el2009urban, boulieri2017space}, whereas the network lattice strategy gained popularity during the last years thanks to increasing computing capabilities and the rapid development of open-source spatial databases (such as Open Street Map) that provided the starting point for creating street networks at a wide spatial scales \parencite{barua2014full, barrington2017world, ma2017multivariate, briz2021modeling, Gilardi2022Multivariate}. Both frameworks employ spatial smoothing techniques to simplify the estimation process, borrowing strength from neighbouring sites. However, in this paper we adopt the network lattice approach because it provides spatially disaggregated results at the street segment level that can be more informative from a social and policy perspective. Furthermore, the road infrastructure and the traffic volumes, which are key ingredients for a road safety model, can be included more naturally in statistical models developed over a network lattice. We refer to \textcite{lord2010statistical}, \textcite{savolainen2011statistical}, and \textcite{ziakopoulos2020review} for exhaustive historical reviews, descriptions of alternative modelling strategies, and additional considerations.

Following the development of statistical methodologies, the road safety analysts focused on studying the car crash determinants. In fact, there exists a vast literature that links road accidents to a variety of factors such as vehicles characteristics \parencite{Lie2006vehicles}, environmental conditions \parencite{Shefer1997environment, Andreescu1998environment}, drivers behaviour \parencite{Horwood2000drinking}, and, as already mentioned, the road design \parencite{milton1998relationship}. Amongst the various potential causes, traffic flows are certainly of particular interest and a solid understanding of their relationship with road casualties is key to improving traffic conditions and reducing crash frequencies. A number of papers (see, for example, \textcite{Wagner2021CarFerw&Flow}) addressed this issue, typically reporting that higher traffic flows are associated with a higher number of collisions \parencite{chang2005analysis, imprialou2016re}. More recently, \textcite{papadimitriou2019review} analyse a series of factors (traffic flows, road type, road surface, \dots) and develop an index to rank their importance in a road safety context. From their conclusions, it emerges that traffic volumes are correlated with higher crash counts (which usually exhibit lower severity levels in case of high congestion). 

Road traffic is a dynamic phenomenon evolving both in space and time. It is typically difficult to obtain precise measurements of traffic volumes for an entire road network and, for this reason, several authors adopted alternative proxies derived from regression models, travel surveys, network characteristics, or origin-destination tables \parencite{LOWRY201498, pun2019multiple, Gilardi2022Multivariate}. Nevertheless, there is increasing evidence suggesting that traffic conditions can be accurately estimated using vehicular GPS data, i.e. moving devices acting as sensors \parencite{Este1999GPS, el2011GPSdata, woodard2017GPSpredicting}. These new technologies represent a cheap and easy-to-use alternative to expensive and time-consuming questionnaires. Moreover, they ease the collection process even at a very granular spatial resolution, creating new opportunities to investigate in depth the relationship between traffic flows and road casualties. For example, \textcite{Stipancic2017GPS} and \textcite{petraki2020combined} used GPS data to correlate collision severities/frequencies with quantitative measures of congestion derived from smartphone-collected GPS data. Following this approach, in this paper we approximate traffic volumes at the street segment level using estimates derived from TomTom Move service \parencite{tomtommove}. The main advantage of our workflow over the previous proposals is that TomTom collects data with global coverage using billion of GPS devices. Therefore, it provides better approximations of local traffic flows than any ad-hoc application.

Nevertheless, spatial data are often prone to measurement error (ME) and traffic figures derived from GPS devices represent no exception. ME can arise at different stages of the data collection process and is typically linked to various sources such as: a) instrumental imprecision in the measurement of physical attributes; b) alignment and harmonisation of characteristics recorded at different spatial scales or domains; c) unobservable effects that are only approximated by surrogate information; d) preferential sampling or incomplete observations \parencite{carroll2006measurement}. Considering the context analysed in this paper, the traffic volumes computed from GPS devices may suffer from ME since, usually, only a certain fraction of the vehicles circulating on a road network is equipped with a (TomTom) GPS receiver. Therefore, the estimates could suffer from underreporting which may not be homogeneous in all parts of the network (i.e. it might have a spatial structure). 

It is well known that ignoring ME leads to severe distortions in statistical inference. The estimate of the regression coefficient associated with the imprecise covariate can be biased downwards (attenuation) or upwards (reverse attenuation), and even the effects of  error-free regressors can be distorted, where the direction of the bias depends on the correlations between the variables. Furthermore, ME may cause a loss of power for detecting signals, whereas relevant features in the data can also be masked. \textcite{carroll2006measurement} call these effects the \textit{``Triple Whammy of Measurement Error"}. However, there are only a few studies that addressed this problem, adjusting for measurement error in spatial modelling. For example, the seminal papers by \textcite{BERNADINELLIetal1997} and \textcite{Xia1998SpatiotemporalMW} introduced Bayesian spatial and spatio-temporal models for disease mapping with errors in the covariates. \textcite{Li2009SpatialLM} developed spatial linear mixed models for a covariate observed with error, whereas \textcite{huque2016} proposed a semi-parametric regression model to correct ME in the spatial explanatory variables. More recently, in the context of road safety analysis, \textcite{XIE2018285} developed a classic measurement-error model to mitigate the potential bias due to measurement errors in traffic volumes when modelling pedestrian crashes at signalised intersections, whereas \textcite{KAMEL2020105612} extended the classic measurement-error model to a spatial setting to account for measurement error in traffic exposure when modelling cyclist-vehicle crashes. Finally, \textcite{XU2022106518} introduced a spatial Bayesian model to address the uncertainty arising from incomplete covariate information (called \emph{exposure data} in their paper)  when modelling bicycle crashes. This is a problem strictly related to the one addressed here since augmenting the data to adjust for incomplete observation results in measurement error in the spatial variable.

Given the severe implications of ME, the objective of this paper is to define a statistical model to estimate the car crashes rate at a very detailed spatial resolution (i.e. the street segments) using extensive information on traffic volumes derived from GPS data and adjusting the statistical model to account for a spatial ME component defined over a network lattice. Furthermore, the analysis presented hereinafter aims at showing how modern data-collecting technologies can be usefully integrated into the statistical analysis of safety mobility. Finally, we improve upon the previous works by suggesting how extremely precise, albeit spatially sparse, instrumental information on traffic flows (such as those obtained from automatic traffic counters) can be used to quantify the ME variability and justify with great care the specification of the ME elements. The suggested methodology is exemplified by analysing the car accidents that occurred in the road network of Leeds (UK) from 2011 to 2019.

We focus on a Bayesian hierarchical approach where prior knowledge can be easily incorporated into the model. The estimation process is worked out using the Integrated Nested Laplace Approximation (INLA), which is an alternative to MCMC inference for the class of latent Gaussian models \parencite{Rue2009, INLA1}. Avoiding sampling, the INLA methodology can be used for efficient model fitting even in presence of large datasets and it makes prior sensitivity analysis and model comparisons more feasible. Similarly to \textcite{muff2015bayesian}, we adjust our model definition using a reformulation of the hierarchy with augmented pseudo-observations. 

The rest of the paper is structured as follows. In Section~\ref{sec:data}, the different data sources are presented, as well as the methods adopted to integrate them into a unique dataset. In Section~\ref{sec:methods}, we introduce the modelling strategy and the classical and spatial frameworks for ME. Section~\ref{sec:results} shows the main results of our analysis, whereas discussion and conclusions in Section~\ref{sec:conclusion} end the paper.

\section{Data}
\label{sec:data}

\begin{figure}
    \centering
    \includegraphics[width = \linewidth]{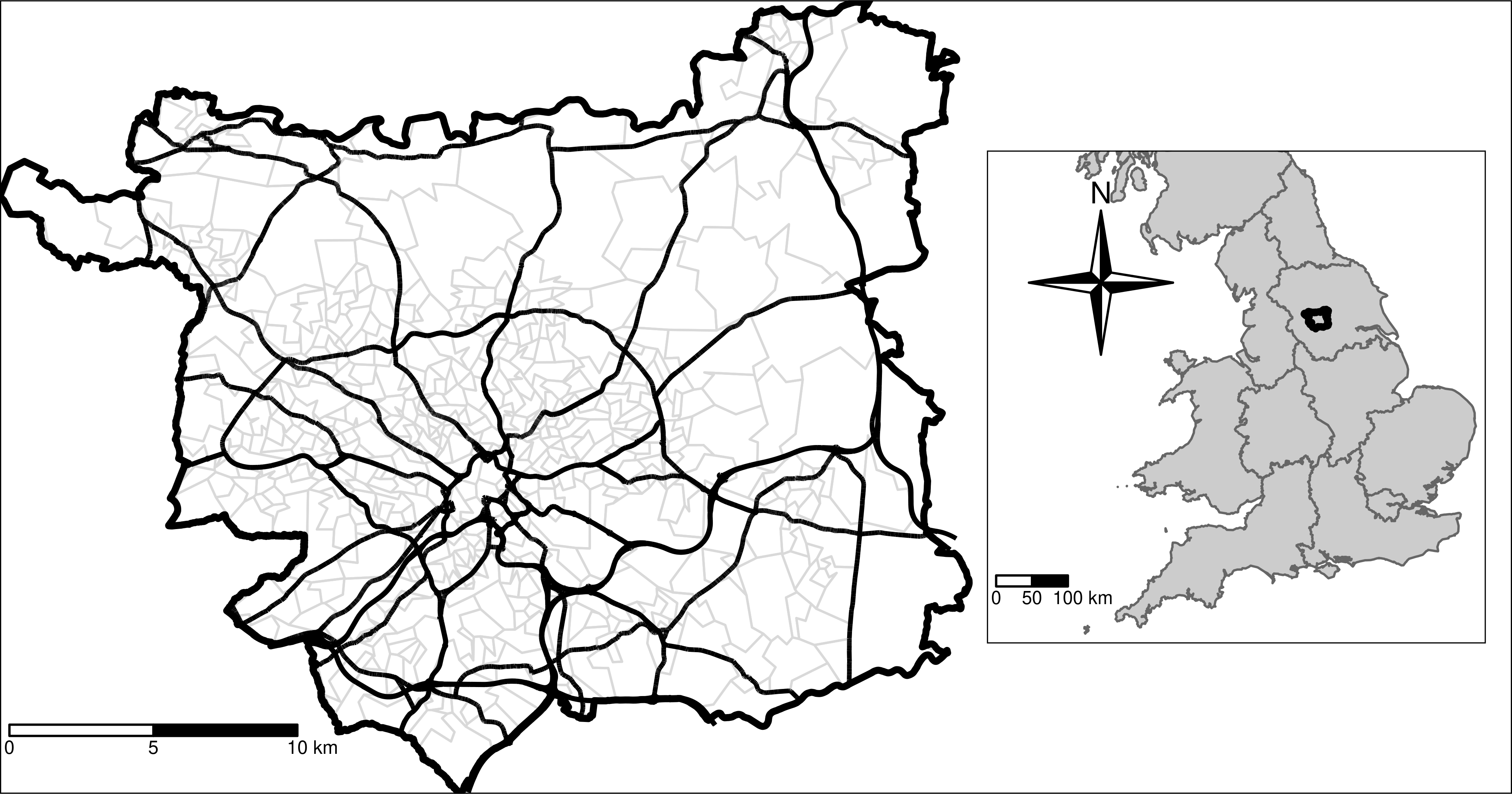}
    \caption{The black polygon denotes the geographical border of the City of Leeds (UK), while the grey polygons denote the Lower Layer Super Output Areas (LSOA) in Leeds. The black segments represent the street network adopted in this study, whereas the inset map is used to locate the study-area with respect to England.}
    \label{fig:study_area}
\end{figure}

As already mentioned, in this paper we analyse the car crashes that occurred in the road network of Leeds from 2011 to 2019. The city of Leeds was selected because it is the most important city in the West Yorkshire region and it accounts for approximately 40\% of all car crashes in the area. The road network and the traffic volumes, which respectively represent the spatial domain of the problem and the covariate suffering from measurement error, were obtained from TomTom Move service\footnote{URL: \url{https://move.tomtom.com/}. Data downloaded in July 2021.}, a \textit{``self-service product that gives direct access to the industry’s largest historical traffic database"} \parencite{tomtommove}. The study area and the road network are depicted in Figure~\ref{fig:study_area}, where the inset map is used to geolocate the city with respect to England. To quantify the variability of the ME process, we considered a second set of traffic counts that was downloaded from the webpage of the Department for Transport (DfT). These figures represent an accurate estimate of the Average Annual Daily Flow (AADF) which is available for approximately 200 sites spread over several parts of the municipality. Finally, the regression models include a set of socio-economic, demographic, and (road) structural variables obtained from the 2011 UK Census and Open Street Map servers, respectively.

The datasets considered in this paper are provided by a range of different agencies and institutions and their harmonisation required some preprocessing. Therefore, in the remaining part of this section we briefly introduce the data providers and describe the steps that were necessary to combine these datasets into a unique structure suitable for the statistical analysis detailed in Section~\ref{sec:methods}.

\subsection{Road network and traffic volumes}
\label{sec:data-network}

The road network was built using data downloaded from TomTom Move. Starting from 2008, TomTom collects anonymised GPS location data and gives  its users access to \textit{``the largest car-centric traffic database of more than 14 trillion anonymously collected real trip data points"}\footnote{Sources: \url{https://support.move.tomtom.com/ts-introduction/} and \url{https://support.move.tomtom.com/products/traffic-stats/}. Last access: November 2022.}. 
The TomTom Traffic Stats service (which is part of TomTom Move) can be used to download traffic data, either via a web portal or an API, using customised queries to select particular geographical areas (e.g. Leeds) and time periods. It offers three types of analyses named \textit{Route Analysis} (which returns average speed and traffic counts for a given route), \textit{Area Analysis} (which returns average speeds, travel times, and traffic counts for all street segments in a given area), and \textit{Traffic Density} (which is similar to \textit{Area Analysis} but focuses only on traffic counts). In all cases, the figures are supported by a geographical database that describes the spatial dimension of the query, typically as a collection of geo-located segments. 

Hence, using the \textit{Area Analysis} service, we downloaded the street segments that compose the most important roads in Leeds and the corresponding traffic volumes. More precisely, TomTom developed a set of rules to rank all segments in a road network according to their importance in a transportation system using a value (named \textit{Functional Road Class}, FRC) going from 0 (Motorways) to 8 (Least important roads). We focused our analysis on a subset of segments, selecting only those that are internally classified by TomTom as \textit{Motorways} (FRC = 0), \textit{Major Roads} (FRC = 2) or \textit{Secondary Roads} (FRC = 3). The chosen segments represent only a subset of the complete city network but, according to our exploratory analysis, approximately 50\% of all car crashes registered in Leeds from 2011 to 2019 occurred in their proximity. Additional details on the internal road classes defined by TomTom are reported at the following link: \url{https://developer.tomtom.com/traffic-stats/documentation/product-information/faq} (Last access: November 2022).

The road network downloaded from TomTom Traffic Stats was composed of approximately 12000 segments of varying lengths covering approximately 600km. According to TomTom documentation, these segments are created internally at every location where a road attribute (e.g. road class or speed limit) changes. After downloading the raw geographic database and adjusting some inconsistencies in the data, we computed a first-order (sparse) binary adjacency matrix among the observations considering a binary relationship where two units are assumed to be neighbours if they share a point in the union of their geographical boundaries. This type of spatial predicate guarantees we don't create fictitious links among roads lying at different levels (e.g. bridges and underpasses). This adjacency matrix, which describes the lattice connectivity and characterises the spatial dimension of the data, is one of the key ingredients to estimate the spatial random effects presented in Section~\ref{sec:methods}. The road network and the FRC values are depicted in Figure~\ref{fig:tomtom_crashes}. We notice that the spatial network spreads uniformly over the entire municipality territory. Moreover, the shape of the motorways and the arterial thoroughfares reaching the city centre can also be clearly distinguished. 

The most important benefit of road networks downloaded from TomTom Traffic Stats is that the provider links the geographic data to traffic counts which, according to the official documentation, are estimated at the segment level using \textit{``anonymous signals from connected car in-dash navigation systems, portable navigation devices and anonymous GPS-equipped mobile phones"}. The information is collected from \textit{``600 million devices in use globally, generating over 3.5 billion kilometres of GPS measurements every single day in over 80 countries"} \footnote{Source: \url{https://support.move.tomtom.com/faq-data-source-quality/}}. Clearly, the estimates obtained from TomTom Move suffer some underreporting since only a proportion of the vehicles circulating on a road network is equipped with a TomTom device. Hence, this variable will be included in the road safety models using a measurement error correction, as explained in Section~\ref{sec:methods}. 

Figure~\ref{fig:tomtom_crashes_counts} displays the estimates of the traffic volumes in the road network of Leeds from the 1st of January 2019 to the 31st of December 2019. First, we notice that the traffic counts are provided at an extremely detailed spatial resolution that let us define a measurement error model at the network lattice level. We can also see that the values have a large variability, ranging from 180 to, approximately, 4.5M units. Unsurprisingly, the highest values are recorded for motorways and major roads, indicating a high correlation level between the two variables (i.e. FRC classes and TomTom counts). Finally, we point out that due to traffic data availability, we developed our analysis considering only one year of historical data. However, it should be considered that annual traffic flows are reasonably stable; hence we do not expect any major difference in the model results if traffic values were aggregated over a larger period \parencite{boulieri2017space}.

\subsection{Car crashes data}

\begin{figure}
    \centering
    \subfloat[][Street segments and car crashes locations. The black dots denote the location of the car crashes. \label{fig:tomtom_crashes}]{
    \includegraphics[width=0.45\linewidth]{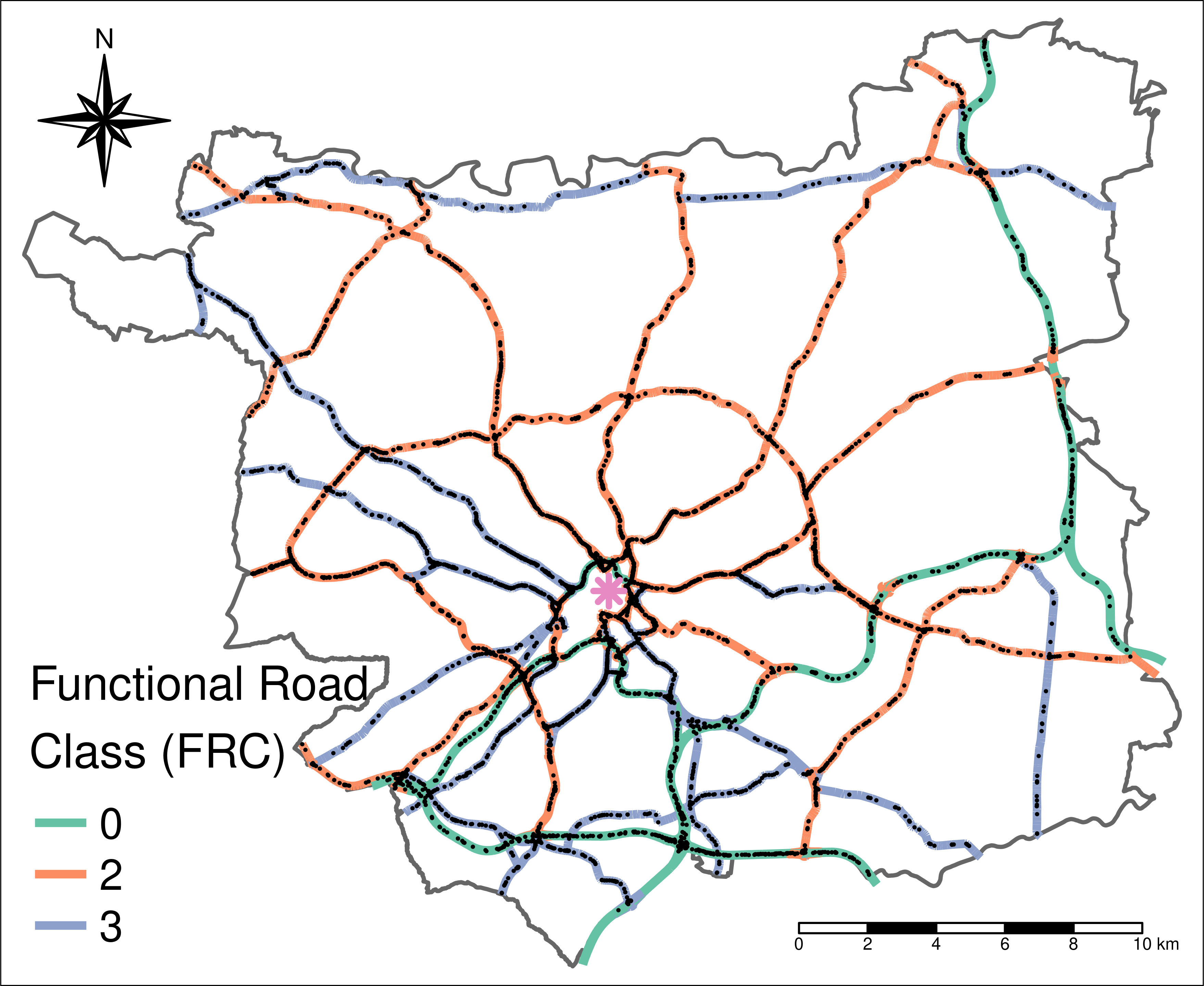}
    }
    \quad
    \subfloat[][Traffic counts measured by TomTom in Leeds during 2019 \label{fig:tomtom_crashes_counts}]{
    \includegraphics[width=0.45\linewidth]{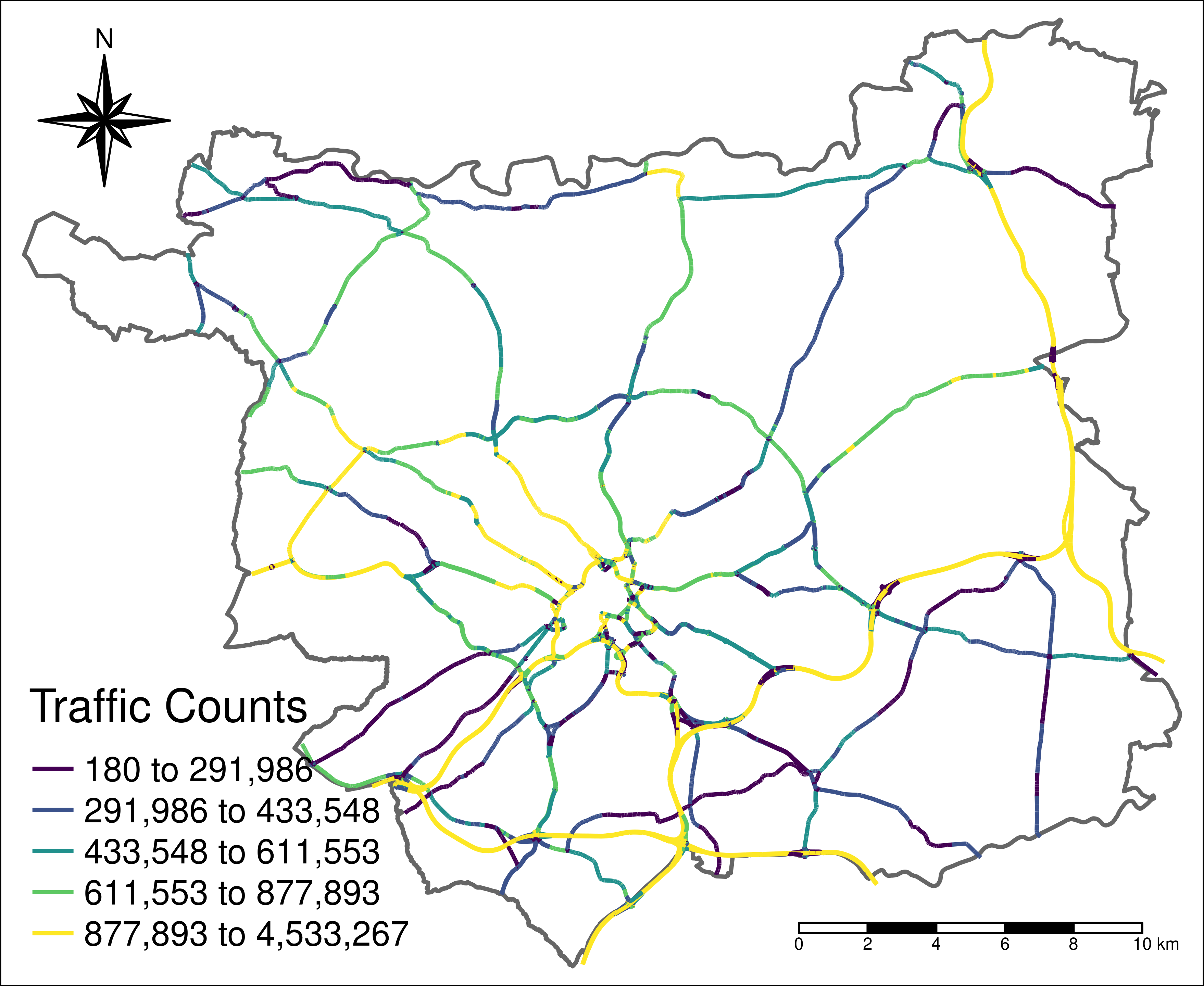}
    }
    \caption{(a) The FRC values describe the importance of each segment in a transportation system. The value 0 corresponds to motorways, 2 is for major roads and 3 for secondary roads. The pink star denotes the city centre. (b) Choropleth map displaying the traffic volumes estimates.}
    \label{fig:crashes}
\end{figure}

We focused on the car crashes that occurred between the 1st of January 2011 and the 31st of December 2019 in the road network of Leeds. More precisely, starting from a database\footnote{Data available at the following url: \url{https://data.gov.uk/dataset/cb7ae6f0-4be6-4935-9277-47e5ce24a11f/road-safety-data}. Downloaded in November 2021.} shared by the DfT that contains all geo-located traffic collisions that occurred in England during the last years, we filtered only the events that happened inside the polygon of Leeds. Then, considering that the TomTom network represents only a subset of the complete city network and that the car crashes locations are typically provided at 10m or less resolution \parencite{dft2011, dft2018}, we excluded all road collisions that occurred farther than 10m from the closest street segments since we assumed they occurred in other segments not included in the network. The final sample is composed of 7125 events which are reported as black dots in Figure~\ref{fig:tomtom_crashes}. Finally, we projected all crash locations to the nearest point of the road network, counting the occurrences on all street segments. These values represent the response variable for the statistical models defined below. 

We end this section by pointing out that the DfT database contains only road collisions that involved at least one personal injury and became known to the Police forces within thirty days of the occurrence. Moreover, the English Government does not enforce people to report all personal injury accidents to the police. Hence, we acknowledge that the crash counts may suffer, to some extent, from under-reporting and we refer to \textcite{savolainen2011statistical} for an extensive discussion on this problem. 

\subsection{Complementary traffic estimates}
\label{sec:datacomplementary}

\begin{figure}
    \centering
    \subfloat[][Locations of the \textit{count points} in Leeds. \label{fig:dftlocationss}]{
    \includegraphics[width=0.45\linewidth]{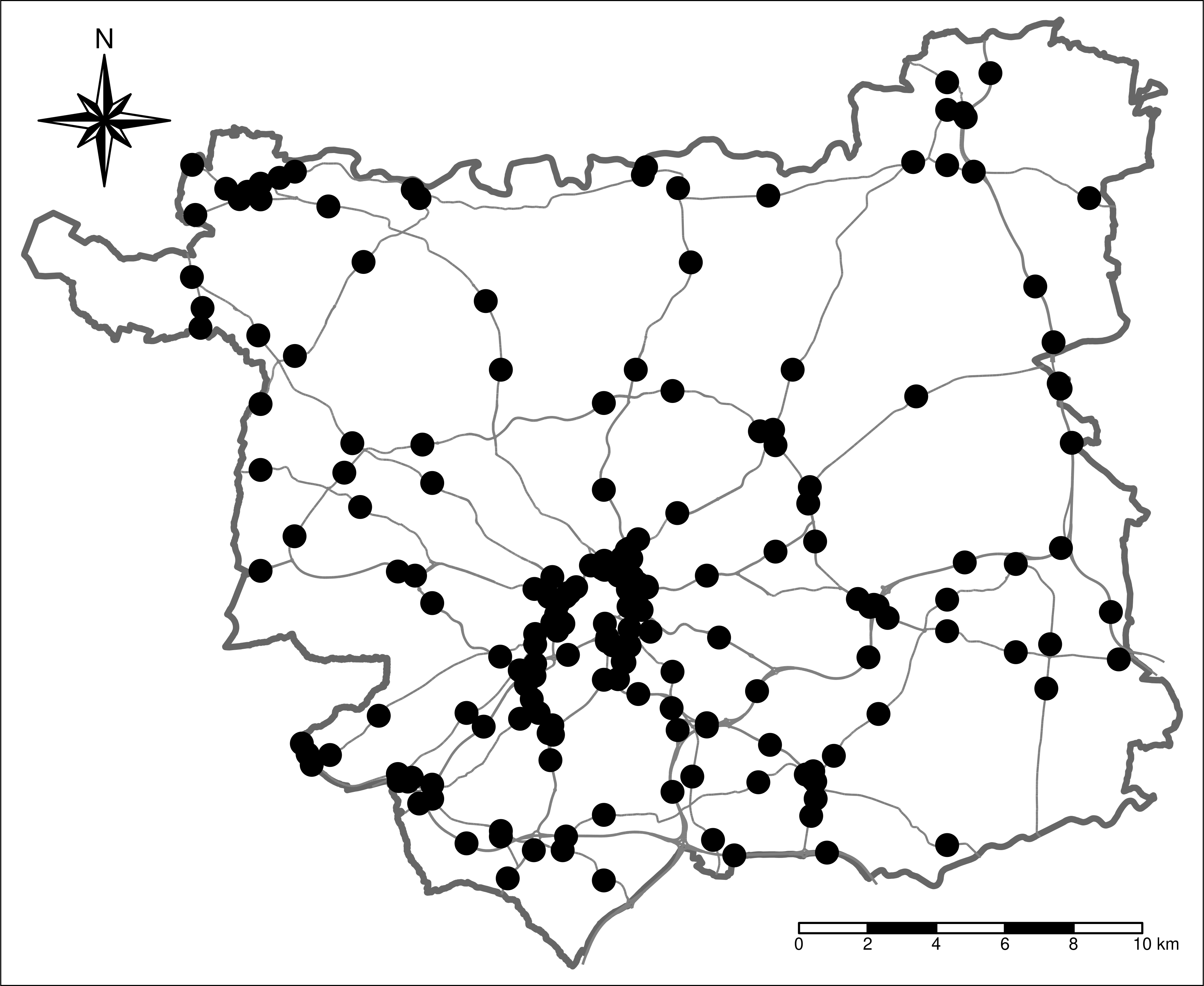}
    }
    \quad
    \subfloat[][Scatterplot displaying the two sets of traffic estimates (log-scale). \label{fig:tomtomanddft}]{
    \includegraphics[width=0.45\linewidth]{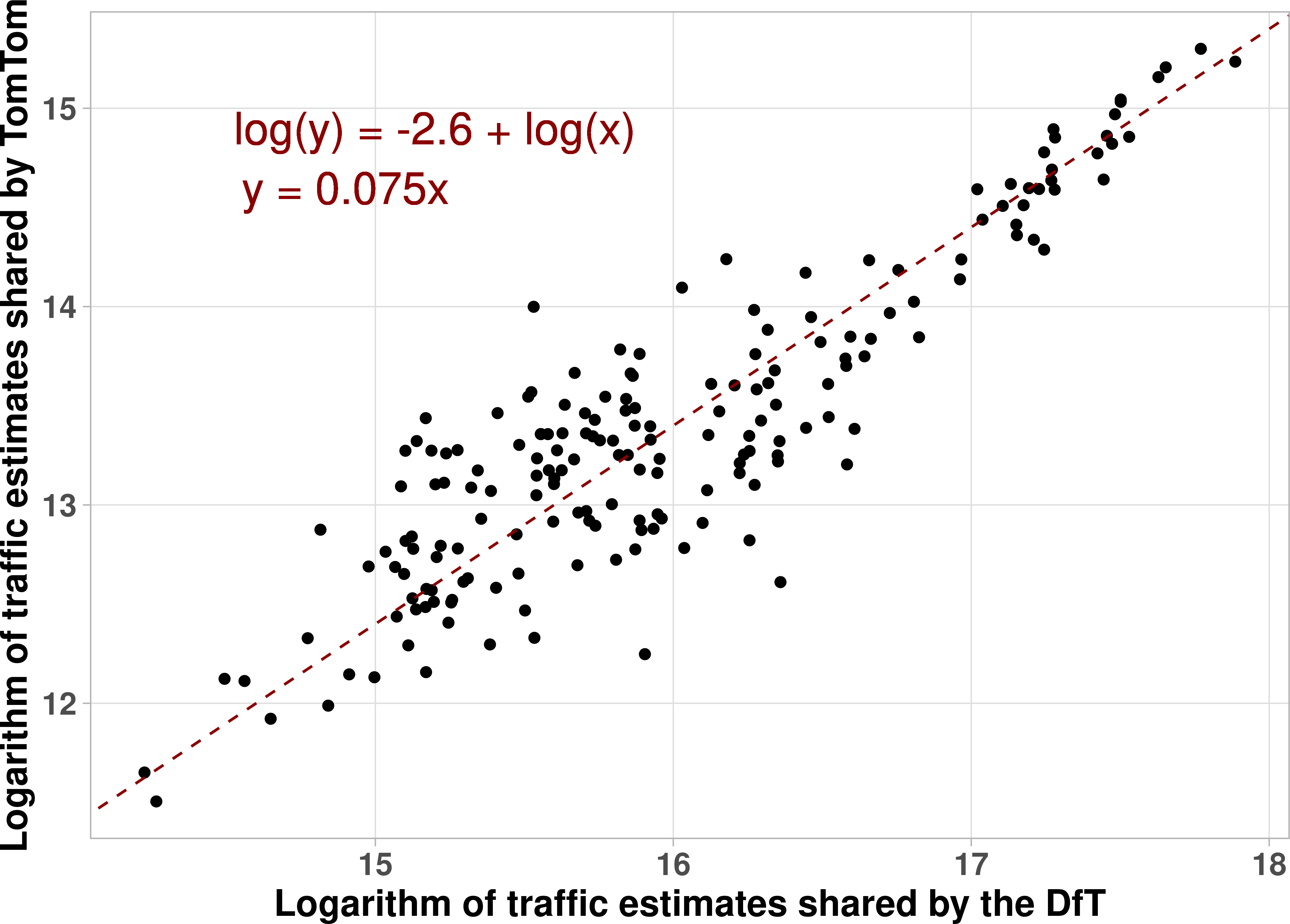}
    }
    \caption{(a) Descriptive analysis of the complementary traffic data. The dashed line displayed in Figure (b) represents the least-square regression line.}
    \label{fig:secondsetofcounts}
\end{figure}

The DfT publishes annual series of road traffic estimates that are based on automatic traffic counters (ATC) and manual counts carried out by trained enumerators at a number of sites named \textit{count points} \parencite{trafficmethodology}. Traffic measurements provided by the DfT do not suffer underreporting as they take into account all types of vehicles travelling at the count points except pedal cycles \parencite{trafficmethodology2}. Hence, we argue that they represent the perfect complementary dataset to calibrate the variability of the terms included in the specification of the ME models detailed in the next Section. Therefore, after downloading the relevant dataset from the DfT web-page\footnote{URL: \url{https://roadtraffic.dft.gov.uk/downloads}. Data downloaded in September 2022.}, we retained only the observations collected during 2019 that pertain to the network of Leeds. The selected count points are displayed in  Figure~\ref{fig:dftlocationss} and we can clearly notice that, despite being more spatially sparse than the Tomtom data, they are uniformly spread over the municipality.

We explored the relationship between the two sets of traffic counts (i.e. GPS- and ATC-derived) considering all road segments where both types of information are available. Figure~\ref{fig:tomtomanddft} shows the scatterplot of the relevant values in log-scale along with the least-square regression line and its equation. In particular, the regression has been estimated by constraining the coefficient associated with the logarithmic transformation of the DfT traffic counts to one and this amounts to assuming that the regression on the original scale passes through the origin, then giving an estimate of the underreporting of mobile device data as a proportionality coefficient between the two quantities. The scatterplot highlights two key aspects which will be incorporated in the definition of the Bayesian hierarchical models described in the next section: a) the traffic counts collected by TomTom represent, on average, 7.5\% of the real flows; b) there exists a good agreement and a linear relationship between the two sets of measurements in log-scale.  

\subsection{Socio-economic characteristics and road infrastructure}
\label{sec:data-census}

The statistical models described in Sections~\ref{sec:methods} and~\ref{sec:results} encompass a set of socio-economic covariates and road infrastructure data obtained from TomTom Move provider, Open Street Map (OSM) servers, and the 2011 UK census. Considering the road infrastructure data, we included the FRC variable described in Section~\ref{sec:data-network} and, after downloading the relevant information from OSM servers, we defined two binary variables that, respectively, record the presence of a traffic light or a pedestrian crossing in each segment of the network.

The socio-economic variables were downloaded from the Nomis website\footnote{URL: \url{www.nomisweb.co.uk}. Data downloaded in October 2022.}, a webpage maintained by the University of Durham which provides all the official statistics related to census and labour market. In particular, following the literature on road safety analysis, we decided to include five variables, namely the \textit{population density} (given by the ratio of the number of inhabitants in a given region and its area in squared metres), the \textit{proportion of young people} (given by the ratio of the people aged 18 to 29 and the total population), \textit{proportion of home workers} (given by the ratio of home workers population and total workers population), the \textit{proportion of working population} (given by the ratio of working and total population in a given area), and the \textit{proportion of households with four or more cars or vans}.

However, the census variables obtained from Nomis webpage were recorded at the Lower Layer Super Output Areas (LSOA) level, i.e. geographical areas designed by the Office for National Statistics to improve the reporting of small area statistics and census data estimates \parencite{guide_UK_geography}. Therefore, these socio-economic covariates and the road network are spatially misaligned and they were merged using an overlay operation: we assigned to each street segment the census values of the LSOA that intersects the largest fraction of the segments. The road network and the LSOA in Leeds are depicted in Figure~\ref{fig:study_area}. 

These covariates measure socio-demographic factors that cannot be fully explained using the road-specific data detailed before. Table~\ref{tab:tab1} summarises a set of descriptive statistics for all the relevant variables included in the regression models. 

\begin{table}
    \centering
    \caption{Descriptive analysis of all variables included in the statistical models. The dashed lines identify three groups of variables, namely the response variable, the socio-economic covariates, and the road characteristics.}
    \label{tab:tab1}
    \begin{tabular}{>{\raggedright\arraybackslash}p{0.2875\textwidth}rrrrc}
      \toprule 
      & \multicolumn{5}{c}{\textbf{Summary Statistics}} \\ 
      \cmidrule{2-6} & {Mean/Freq.} & {Std.Dev} & {Min.} & {Max.} & {Histogram} \\
      \midrule \textbf{Car crashes counts} & 0.589 & 1.220 & 0 & 19 &
      \raisebox{-0.2\totalheight}{\includegraphics[width=0.15\linewidth]{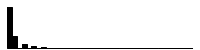}} \\
      \cdashlinelr{1-6}
      \textbf{Population density} & 23.051 & 23.246 & 0.5 & 157.1 & \raisebox{-0.2\totalheight}{\includegraphics[width=0.15\linewidth]{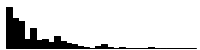}} \\
      \textbf{Young residents (\%)} & 18.495 & 15.075 & 5.482 & 86.641 & \raisebox{-0.2\totalheight}{\includegraphics[width=0.15\linewidth]{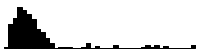}} \\ 
      \textbf{People working from home (\%)} &
      \raisebox{-\totalheight}{3.299} & \raisebox{-\totalheight}{2.025} & \raisebox{-\totalheight}{0.270} & \raisebox{-\totalheight}{9.421} & \raisebox{-0.66\totalheight}{\includegraphics[width=0.15\linewidth]{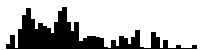}} \\
      \textbf{Working pop. (\%)} & 48.354 & 8.707 & 25.977 & 84.366 & \raisebox{-0.2\totalheight}{\includegraphics[width=0.15\linewidth]{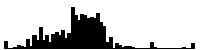}} \\
      \textbf{Households with 4+ vehicles (\%)} &  \raisebox{-\totalheight}{0.015} & \raisebox{-\totalheight}{0.014} & \raisebox{-\totalheight}{0} & \raisebox{-\totalheight}{0.097} & \raisebox{-0.66\totalheight}{\includegraphics[width=0.15\linewidth]{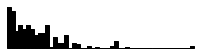}} \\
      \cdashlinelr{1-6}
      \addlinespace
      \textbf{Presence of pedestrian crossings} & \raisebox{-.66\totalheight}{0.092} &
      \raisebox{-.66\totalheight}{-} & \raisebox{-.66\totalheight}{-} & \raisebox{-.66\totalheight}{-} \\
      \textbf{Segments' lengths} & 49.031 & 84.886 & 0.657 & 2756 & \raisebox{-0.2\totalheight}{\includegraphics[width=0.15\linewidth]{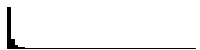}}\\
      \textbf{Traffic Volumes} & $7.0 \times 10^5$ & $6.5 \times 10^5$ & 180 & $4.3 \times 10^6$ & \raisebox{-0.2\totalheight}{\includegraphics[width=0.15\linewidth]{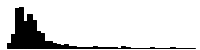}} \\
      \textbf{Presence of traffic signals} & \raisebox{-\totalheight}{0.073} &
      \raisebox{-\totalheight}{-} & \raisebox{-\totalheight}{-} & \raisebox{-\totalheight}{-} \\
      \textbf{Functional Road Class (FRC):} \\ 
      \hspace{1em} 0 (Motorways) & 0.091 & & & & \multirow{3}*{\raisebox{-0.01\totalheight}{\includegraphics[width=0.15\linewidth]{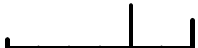}}}\\
      \hspace{1em} 2 (Major roads)& 0.555\\
      \hspace{1em} 3 (Secondary roads) & 0.354 \\
      \bottomrule
    \end{tabular}
\end{table}

\section{Statistical methods}
\label{sec:methods}

This section introduces the statistical methods that were developed to analyse the road casualty occurrences. We start from a three-level Bayesian hierarchical model which is presented in Section~\ref{sec:baseline-model}. This model does not include any measurement error correction and represents the baseline for our analyses. Thereafter, Sections~\ref{sec:classicalME} and~\ref{sec:spatialME} define two types of extensions. The first one enhances the baseline including a (classical) ME term, whereas the second one further improves the correction allowing for potential spatial dependence in the ME by including a spatially structured random effect. Finally, we briefly discuss a few techniques for comparing the three approaches and some computational details behind the estimation of ME models. 

\subsection{Baseline model}
\label{sec:baseline-model}

Let $y_i, \ i = 1, \dots, n$ denote the number of car crashes that occurred in the $i$th road segment. Following a classical hypothesis in the road safety literature (see, for example, \textcite[Section~3]{ziakopoulos2020review}), in the first stage of the hierarchy we assume that 
\begin{equation}
y_{i}|\lambda_{i} \sim \text{Poisson}(e_i\lambda_{i}),
\label{eq:first-level-baseline}
\end{equation} 
where $\lambda_{i}$ represents the car crashes rate and $e_i$ is an offset parameter that we set equal to the geographical length of each segment. As we mentioned in the previous section, the road segments have different lengths; hence, the offset values account for the fact that longer street segments are expected to have a higher collision risk than shorter ones, guaranteeing comparable rates. 

The second level of the hierarchy defines a log-linear structure for $\lambda_{i}$. More precisely, we assume that  
\begin{equation}
\log(\lambda_{i}) = \beta_0 + \beta_x  \log(x_i) + \sum_{j = 1}^{p}\beta_{j}z_{ij} + \theta_i,
\label{eq:second-level-baseline}
\end{equation}
where $\beta_0$ denotes the intercept, $x_i$ represents the unobserved true traffic volumes with regression coefficient $\beta_x$, $\lbrace z_{i1}, \dots, z_{ip}\rbrace$ is a set of error-free covariates and $\lbrace \beta_1, \dots, \beta_p\rbrace$ are the corresponding coefficients. Following the ideas in \textcite[Section 2.1]{XU2022106518}, we  included the traffic covariate $x$ after taking a logarithm transformation considering its non-negative, continuous, and positively skewed nature, which is also displayed in Table~\ref{tab:tab1}. Clearly, model~\eqref{eq:second-level-baseline} cannot be estimated since $x_i$ is not observed. Therefore, we assume that the traffic estimates obtained from TomTom Move represent a surrogate, say $w_i$, for the real traffic volumes and, after substituting $x_i$ with $w_i$, the log-linear structure for $\lambda_i$ in the baseline model writes 
\begin{equation}
\log(\lambda_{i}) = \beta_0 + \beta_x \log(w_i) + \sum_{j = 1}^{p}\beta_{j}z_{ij} + \theta_i.
\label{eq:second-level-baseline-est}
\end{equation}
Finally, $\theta_i$ denotes a spatially structured random effect that is modelled using an Intrinsic Conditional Auto-Regressive (ICAR) distribution \parencite{besag1974spatial, besag1995conditional}.

The ICAR distribution is a common tool to induce spatial dependence in statistical models developed for lattice data. It is typically defined through a set of conditional distributions
\begin{equation}
\theta_i | \left\lbrace \theta_{i'}, i' \in \partial_i\right\rbrace; \tau_\theta \sim N\left(\frac{1}{m_i}\sum_{i'\in\partial_i}\theta_{i'}; \frac{1}{m_i\tau_\theta}\right), 
\label{eq:ICAR-conditional}
\end{equation}
where $m_i$ and $\partial_i$ represent the cardinality and the indices of the set of neighbours for unit $i$, respectively. As we briefly mentioned in Section~\ref{sec:data-network}, these quantities are derived from a (sparse) binary adjacency matrix, here denoted by $\bm{W}$, that summarises the neighbouring structures among the segments. 
\textcite{besag1974spatial} showed that the conditional distributions in~\eqref{eq:ICAR-conditional} yield a joint multivariate distribution that can be expressed as
\begin{equation}
\bm{\theta}|\tau_\theta \sim N\left(\bm{0}; [\tau_\theta(\bm{D} - \bm{W})]^{-1}\right), 
\label{eq:ICAR-joint}
\end{equation}
where $\bm{D} = \text{diag}(m_1, \dots, m_n)$. It is possible to prove that the variance-covariance matrix in Equation~\eqref{eq:ICAR-joint} is not positive definite, a problem that is usually solved by imposing a sum-to-zero constraint on each component (i.e. each cluster of connected road segments) of the vector $\lbrace \theta_1, \dots, \theta_{n}\rbrace$ \parencite{hodges2003precision}. We refer to \textcite{banerjee2015hierarchical}, \textcite{martinez2019disease}, and references therein for more details on the ICAR prior.  

The third level completes the definition of the baseline hierarchical model specifying the prior distributions for each parameter in equation~\eqref{eq:second-level-baseline}. In this case, we assigned an uninformative $N(0, 50)$ prior to $\beta_0$, $\beta_x$, and $\beta_{j}, \ j = 1, \dots, p$ and a logGamma prior to $\log(\tau_\theta)$ with parameters $1$ (shape) and $5e-05$ (rate). 

\subsection{Classical measurement error model}
\label{sec:classicalME}

As we mentioned at the beginning of this section, the first extension (hereby denoted as \textit{Extension I}) improves over the baseline model by adjusting the estimates of the traffic volumes provided by TomTom with a \textit{classical} ME correction. In particular, considering that, as described in Section~\ref{sec:data}, the surrogate variable $w$ always underestimates the real flows $x$, we decided to adopt a multiplicative ME model
\begin{equation}
w_i = \rho_0 x_i u_i, \quad i = 1, \dots, n,
\label{eq:multiplicativeME}
\end{equation}
where $\rho_0 \in [0, 1]$ is a proportionality parameter that represents the percentage of traffic flow captured by TomTom GPS devices and $\lbrace u_1, \dots, u_n \rbrace$ is a set of random errors assumed to be mutually independent and log-Normally distributed. Furthermore, following the classical theory of ME models \parencite{carroll2006measurement}, we also assume that the error terms are independent of the true covariate and, when working in a regression context, of the response variable and any other explanatory variable.

In log-scale, the multiplicative ME model defined in Equation~\eqref{eq:multiplicativeME} reads 
\begin{equation}
\log(w_i) = \tilde{\rho}_0 + \log(x_i) + \log(u_i),
\label{eq:linearME}
\end{equation}
where $\tilde{\rho}_0 = \log(\rho_0)$ and $\lbrace \log(u_i) \rbrace_{i = 1} ^ n$ denotes a collection of independent and identically distributed Gaussian random variables which are assumed to have zero mean and precision $\tau_u$. Looking at Figure \ref{fig:tomtomanddft}, we note that the model specification in Equation~\eqref{eq:linearME} appropriately describes the relationship between actual and GPS-detected traffic flows when they are computed on the log scale. Moreover, we also observe that the residuals of the log-log regression are homoscedastic, implying that the ME is not related to the magnitude of the traffic volumes and strengthening our assumption of independence between ME and the true value of traffic flows. Finally, following the approach detailed in \textcite[Section~4.1]{muff2015bayesian}, we also assume that, on the log-scale, the unobserved traffic covariate on segment $i$ has a Gaussian distribution with precision $\tau_x$ and mean $\mu_{x, i}$ which depends on a set of (error-free) predictors, say $\tilde{z}_{ij}$. 

After introducing the structure of the classical ME, we now present \textit{Extensions I} which enriches the baseline model by including two terms that account for potential errors in the surrogate covariate. The first level of the hierarchy writes 
\begin{equation}
\begin{cases}
y_{i} \mid \lambda_{i} \sim \text{Poisson}(e_i\lambda_{i}) \\ 
\log(x_i) \mid  \mu_{x, i}, \tau_x \sim N\left(\mu_{x, i}, 1 / \tau_x\right) \\
\log(w_i) \mid \mu_{w, i}, \tau_u \sim N\left(\mu_{w, i}, 1 / \tau_u \right)
\end{cases}
\label{eq:first-level-classicalME}
\end{equation}
where the index $i$ ranges from $1$ to $n$. The first expression in~\eqref{eq:first-level-classicalME}, which is usually named \textit{regression} or \textit{outcome} model in the ME literature, can be interpreted as before, whereas the second and third expressions define the ME structure. They are typically referred to as \textit{exposure} and \textit{error} models, respectively. 

In the second level of the hierarchy, we assume that 
\begin{equation}
\begin{cases}
\log(\lambda_{i}) = \beta_0 + \beta_x \log(x_i) + \sum_{j = 1}^{p}\beta_{j}z_{ij} + \theta_i \\ 
\mu_{x, i} = \alpha_0 + \sum_{j = 1} ^ {q} \alpha_j\tilde{z}_{ij} \\
\mu_{w, i} = \tilde{\rho}_0 + \log(x_i)
\end{cases},
\label{eq:second-level-classicalME}
\end{equation}
where the coefficients $\lbrace \beta_x, \beta_0, \beta_1, \dots, \beta_p \rbrace$ and the variables $\lbrace x_i, z_{ij}, \theta_i \rbrace$ are introduced above, the parameter $\alpha_0$ denotes the intercept in the exposure model and $\lbrace \alpha_1, \dots, \alpha_q\rbrace$ is the set of regression coefficients corresponding to the error-free variables $\lbrace \tilde{z}_{i1}, \dots, \tilde{z}_{iq}\rbrace$. We also note that the ME specification in (7) can be explicitly recovered by combining the ME structure defined in the third equations of (8) and (9). 

The third level of the hierarchy specifies the prior distributions for all parameters included in the regression model. In particular, following the same reasoning as before, we assigned a $N(0, 50)$ prior to $\beta_0$, $\beta_x$, $\lbrace\beta_1, \dots, \beta_p\rbrace$, $\alpha_0$, $\lbrace\alpha_1, \dots, \alpha_q\rbrace$, and a $\text{logGamma}(1, 5e-05)$ prior to the logarithm of $\tau_{\theta}$. The elicitation of the prior distributions for $\tilde{\rho}_0$, $\tau_x$, and $\tau_u$ (i.e. the parameters for the ME analysis) requires more care and it was carried out considering the complementary road traffic data shared by the DfT. In particular, our choices are based on the exploratory analysis regarding the relationship between TomTom estimates and real traffic flows described and displayed in Section~\ref{sec:datacomplementary}. In order to specify the a-priori distribution of $\rho_0$, the proportionality parameter in the multiplicative ME model, we assumed a $\text{log-Normal}(2.6, 0.067)$ prior to the opposite of $\tilde{\rho}_0$ which, in turn, induces an inverse log-log-Normal prior on $\rho_0$ with mean equal to 0.076 and 5-th and 95-th percentile equal to 0.047 and 0.111, respectively. These values were chosen taking into account the slope of the least-square regression line displayed in Figure~\ref{fig:tomtomanddft} and the average penetration rate of TomTom devices on urban highways\footnote{URL: \url{https://devforum.tomtom.com/t/tomtom-traffic-stats-in-scientific-research/2304/3}. Last access on December 2022.} In order to specify the prior distribution of $\tau_x$, we considered again the traffic volumes observed at the count point locations. We estimated the precision of the unobservable traffic covariate in a Bayesian framework and used its posterior distribution as a prior for $\tau_x$. In particular, we adopted a $\text{Gamma}(96.2, 58.64)$ distribution which is centred around 1.62, has the 2.5\% quantile at 1.33 and the 97.5\% quantile at 1.98. Finally, the prior on $\tau_u$ was defined considering the posterior distribution of the unexplained variance in the log-log regression between actual and GPS-detected traffic flows. In particular, using similar considerations as before, we decided to adopt a $\text{Gamma}(96.12, 12.27)$ prior which is centred around 7.8, has the 2.5\% quantile at 6.34, and the 97.5\% quantile at 9.47\%.

\subsection{Spatial classical measurement error model}
\label{sec:spatialME}

The second extension (hereby denoted as \textit{Extension II}) assumes that the measurement error model can also include a spatially structured component which encompasses spatial regularities that are not appropriately accounted for by the measurable covariates. We will refer to this type of structure using the term \textit{spatial classical measurement error model}. 

The classical technique used to induce spatial dependence in a lattice model is the BYM prior \parencite{besag1991bayesian}. Using this approach, the ME model writes
\[
\log(w_i) = \tilde{\rho}_0 + \log(x_i) + u_i + \nu_i,
\]
where $u_i$ represents a zero-mean unstructured Gaussian random error and $\nu_i$ denotes a spatially structured random component typically modelled with an ICAR prior. All the other terms have the same interpretation as before. Unfortunately, there exists a vast literature that discusses the identifiability problems of the hyperparameters in the BYM prior. Therefore, following the suggestions in \textcite{simpson2017penalising}, we decided to adopt a re-parametrisation, hereby denoted as BYM2, that writes 
\begin{equation}
\log(w_i) = \tilde{\rho}_0 + \log(x_i) + \frac{1}{\tau_u}\left(\sqrt{1 - \phi}\tilde{u}_i + \sqrt{\phi}\tilde{\nu}_i\right), 
\label{eq:spatialME}
\end{equation}
where $0 \le \phi \le 1$ is a mixing coefficient that represents the proportion of variance explained by each of the two components while $\tilde{u}_i$ and $\tilde{\nu}_i$ denote a scaled version of the $u_i$ and $\nu_i$ terms having generalised variance equal to one. Finally, the parameter $\tau_u$ describes the marginal precision explained by $\tilde{u}_i$ and $\tilde{\nu}_i$. The main advantage of the BYM2 model over its classical formulation is that the \textit{``hyperparameters $(\phi, \tau_u)$ control different parts of the prior and this naturally allows for independent prior specification"} \parencite{simpson2017penalising}. Given the spatial nature of the data considered in this paper, we believe that spatial ME models provide an ideal framework for appropriately approximating the (unobserved) road traffic volumes at the segment level using data derived from mobile devices.  

Under the aforementioned assumptions, we can finally present the \textit{Extension II} model. The hierarchical structure previously described by Equations from (7) to (9) is now enriched by an extra term ($\phi$) which is included in the ME component following the functional form defined by Equation~\eqref{eq:spatialME}. We further assumed a $\text{Uniform}(0, 1)$ distribution for $\phi$ and a $\text{Gamma}(96.12, 12.27)$ prior for $\tau_u$, since in the BYM2 formulation, the latter term has the same interpretation as the $\tau_u$ hyperparameter included in the classical approach. All the other parameters and the corresponding priors are defined as before.

\subsection{Bayesian estimation of measurement error models}
\label{sec:bayesian-details}

The three models described in the previous sections were estimated using the Integrated Nested Laplace Approximation (INLA), a popular alternative to classical MCMC sampling for a particular class of models, named Latent Gaussian Models (LGM) \parencite{Rue2009}. INLA provides an attractive framework to model spatial dependence in lattice data via random effects that can be conveniently expressed as multivariate Gaussian distributions, typically with a sparse precision matrix \parencite{rue2005gaussian, Rue2009, bakka2018spatial}. The ICAR prior represents a classical example. Moreover, \textcite{muff2015bayesian} recently showed that the INLA approach can be adjusted to accommodate ME terms through a \textit{``reformulation with augmented pseudo-observations and a suitable extension of the latent field"}. Hence, following their recommendations, we modified the last two models presented above slightly readjusting the second level of the hierarchical structure. 

In particular, the exposure model previously defined by
\[
\log(x_i) \mid  \mu_{x, i}, \tau_x \sim N\left(\mu_{x, i}, 1 / \tau_x\right); \quad \mu_{x, i} = \alpha_0 + \sum_{j = 1} ^ {q} \alpha_j\tilde{z}_{ij}
\]
now writes 
\[
\log(x_i) \mid  \mu_{x, i}, \tau_x \sim N\left(\mu_{x, i}, 1 / \tau_x\right); \quad 0 =  -\mu_{x, i} + \alpha_0 + \sum_{j = 1} ^ {q} \alpha_j\tilde{z}_{ij}, 
\]
where we introduced a set of zero pseudo-observations such that the set of response variables can be specified via a response matrix having one column for each separate equation and $3n$ rows. We refer to \textcite{muff2015bayesian} where more details regarding the computing aspects are discussed.

Finally, the models introduced in Sections~\ref{sec:classicalME} and~\ref{sec:spatialME} were compared using Deviance Information Criterion (DIC) \parencite{spiegelhalter2002bayesian} and Watanabe–Akaike Information Criterion (WAIC) \parencite{watanabe2010asymptotic, gelman2014understanding}. These criteria represent a generalisation of classical Akaike information criterion (AIC) to Bayesian hierarchical models. They measure the adequacy of a model penalised by the number of effective parameters. In both cases, lower values of the index suggest a better fit of the model.

\section{Results}
\label{sec:results}

In this section, we summarise the results obtained after estimating the three statistical models detailed before. As already mentioned, given the spatial nature of the problem and the network lattice constraints, we decided to adopt the computationally efficient INLA framework via the homonymous R package \parencite{INLA1, R-software}. Using a laptop with an AMD Ryzen 5 3500U processor with Radeon Vega Mobile Gfx 2.10 GHz, four cores and 8GB of RAM, estimating the baseline model required approximately 45 seconds, whereas estimating \textit{Extensions I} and \textit{II} required approximately 3 and 6 minutes, respectively. 

\subsection{Fixed effects}

\begin{table}
  \centering
  \caption{Posterior means and standard deviations (in brackets) of all fixed effects included in the regression model.}
  \label{tab:fixed-effects-log-lambda}
  \begin{tabular}
  {
  l
  @{}
  S[table-format = +2.3, parse-numbers = false]
  S[table-format = +2.3, parse-numbers = false]
  S[table-format = +2.3, parse-numbers = false]
  @{}
  }
  \toprule
  & {Baseline} & {\textit{Extension I}} & {\textit{Extension II}} \\
  \midrule
  & -7.333 & -12.071 & -11.308 \\
  \multirow{-2}{*}{Intercept} & (0.805) & (1.061) & (0.631) \\
  & 0.457 & 0.799 & 0.743 \\
  \multirow{-2}{*}{FRC 2 (Major roads)} & (0.113) & (0.125) & (0.103) \\
  & 0.483 & 0.964 & 0.923 \\
  \multirow{-2}{*}{FRC 3 (Secondary roads)} & (0.138) & (0.155) & (0.118) \\
  & 0.045 & 0.046 & 0.045 \\ 
  \multirow{-2}{*}{Population Density} & (0.065) & (0.064) & (0.066) \\
  & 0.021 & 0.027 & 0.017 \\ 
  \multirow{-2}{*}{Young residents (\%)} & (0.132) & (0.130) & (0.131) \\
  & 0.116 & 0.128 & 0.128 \\ 
  \multirow{-2}{*}{Work from home (\%)} & (0.126) & (0.124) & (0.126) \\
  & -0.035 & -0.060 & -0.059 \\ 
  \multirow{-2}{*}{Working pop. (\%)} & (0.265) & (0.262) & (0.265) \\
  & 0.149 & 0.162 & 0.159 \\ 
  \multirow{-2}{*}{Households w/ 4+ vehicles} & (0.131) & (0.130) & (0.132) \\
  & 0.309 & 0.314 & 0.311 \\ 
  \multirow{-2}{*}{Presence of pedestrian crossings} & (0.047) & (0.047) & (0.047) \\
  & 0.390 & 0.409 & 0.402 \\ 
  \multirow{-2}{*}{Presence of traffic signals} & (0.050) & (0.050) & (0.050) \\
  & 0.216 & 0.402 & 0.506 \\ 
  \multirow{-2}{*}{Road traffic} & (0.040) & (0.070) & (0.058) \\
  \bottomrule
  \end{tabular}
\end{table}

The baseline model and the two extensions were specified including a set of common covariates in the log-linear structure for $\lambda$. In particular, besides the traffic measurements, we considered the road class indicator (i.e. the FRC variable described in Section~\ref{sec:data-network}) as well as seven socio-economic or structural variables derived from 2011 UK Census and Open Street Map servers. These variables were introduced in Section~\ref{sec:data-census} and their descriptive statistics are summarised in Table~\ref{tab:tab1}. Furthermore, the exploratory analyses highlighted that traffic counts and road types are, to some extent, correlated; hence, this road-specific covariate was also included as a fixed effect in the exposure model (i.e. the second expression in Equations~\eqref{eq:first-level-classicalME} and~\eqref{eq:second-level-classicalME}). 

The posterior means and standard deviations for all regressors included in the observation and exposure models are summarised in Tables~\ref{tab:fixed-effects-log-lambda} and~\ref{tab:fixed-effects-mu}, respectively. The last row in Table~\ref{tab:fixed-effects-log-lambda} shows the estimates associated with the road traffic coefficient. The values highlight the importance of the ME correction and the increasing strength of the attenuation bias which is induced by using an imprecise spatial covariate. The traffic measurements were included in the regression model not linearly; therefore, the figures must be interpreted as unitary increments in the log scale. For example, assuming that the number of annual traffic units increases from $e^{15.9} \simeq 7 \cdot 10 ^ 6$ (the median value registered by the automatic traffic counters for the average street segment) to $e^{16.9}$ (i.e. an increase of approximately 14M vehicles that traverse the average segment in one year) and keeping all the other variables fixed, the logarithm of the car crashes rate increases by 0.216 (baseline), 0.402 (classical ME), and 0.506 (spatial ME). 

\begin{table}
  \centering
  \caption{Posterior means and standard deviations (in brackets) of all fixed effects included in the exposure model.}
  \label{tab:fixed-effects-mu}
  \begin{tabular}
  {
  l
  @{}
  S[table-format = +2.3, parse-numbers = false]
  S[table-format = +2.3, parse-numbers = false]
  @{}
  }
  \toprule
  & {\textit{Extension I}} & {\textit{Extension II}} \\
  \midrule
  & 16.221 & 16.874 \\
  \multirow{-2}{*}{Intercept} & (0.237) & (0.172) \\
  & -1.351 & -1.360 \\
  \multirow{-2}{*}{FRC 2 (Major roads)} $\quad$ & (0.023) & (0.030) \\
  & -1.763 & -2.045 \\
  \multirow{-2}{*}{FRC 3 (Secondary roads)} & (0.024) & (0.033) \\
  \bottomrule
  \end{tabular}
\end{table}

No socio-economic or demographic regressor was found to be significant in the baseline or extension models, whereas the road structural variables proved to be extremely relevant, with motorways (FRC 0) being the safest road type among the three classes (since the coefficients associated with FRC 2 and 3 are always positive). This result is consistent with the road safety literature \parencite{flahaut2004impact, li2008differences, boulieri2017space, US2017, Gilardi2022Multivariate}, and can be explained by taking into account the ad-hoc prevention measures typically implemented by road planners (e.g. physical dividers between opposite directions and absence of road-side obstacles). As one might expect, \textit{pedestrian crossings} and \textit{traffic lights}, which are typically located near high-risk areas on roads of type "2" and "3", were found to increase the risk of car accidents. The magnitude and the sign of the posterior means are generally consistent among the three specifications.

The estimates in Table~\ref{tab:fixed-effects-mu} suggest that, unsurprisingly, motorways are the most congested roads in Leeds. In fact, the coefficients associated with the major and secondary roads were found negative, highlighting that, on average, the corresponding traffic volumes are lower than the reference class (i.e. the motorways). These findings clearly show that defining a proper ME structure is a crucial step for developing an accurate road safety model and mitigating the attenuation bias. 

\subsection{Random effects}

\begin{table}
  \centering
  \caption{Posterior means and standard deviations (in brackets) of all hyperparameters}
  \label{tab:random-effects}
  \begin{tabular}
  {
  l
  @{}
  S[table-format = +2.3, parse-numbers = false]
  S[table-format = +2.3, parse-numbers = false]
  S[table-format = +2.3, parse-numbers = false]
  @{}
  }
  \toprule
  & {Baseline} & {\textit{Extension I}} & {\textit{Extension II}} \\
  \midrule
  & 0.064 & 0.072 & 0.066 \\
  \multirow{-2}{*}{$\tau_\theta\quad$} & (0.004) & (0.004) & (0.004) \\
  & & 2.546 & 4.471 \\
  \multirow{-2}{*}{$\tau_x$} & & (0.075) & (0.133) \\
  & & 0.072 & 0.120 \\
  \multirow{-2}{*}{$\rho_0$} & & (0.020) & (0.024) \\
  & & 7.791 & 1.716 \\
  \multirow{-2}{*}{$\tau_u\quad$} & & (0.612) & (0.105) \\
  & & & 0.905 \\
  \multirow{-2}{*}{$\phi\quad$} & & & (0.005) \\
  \bottomrule
  \end{tabular}
\end{table}

Table~\ref{tab:random-effects} summarises the posterior means and standard deviations of all hyperparameters. The first row contains the estimates of $\tau_\theta$, i.e. the precision of the spatially structured random effect included in the outcome model. This is the only hyperparameter shared by all three specifications. If we compare its values across the different models, we can observe a similar degree of spatial uncertainty. The second row shows the estimates of $\tau_x$, i.e. the residual variability in the linear regression defining the exposure model. The two figures are quite similar, although we observe slightly less variability in the \textit{Extension II} model. The term $\rho_0$ represents the proportion of real traffic observed by TomTom GPS devices. The estimates are highly stable and, in both cases, the proportion is approximately equal to 10\%.

The last two rows in Table~\ref{tab:random-effects} summarise the posterior distribution of the parameters included in the ME model. The first one, namely $\tau_u$, represents the (total) variability of the random error component(s) included in the ME specification. As we can see, \emph{Extension II} exhibits a slightly higher degree of uncertainty compared to the classical approach. Finally, the last coefficient, namely $\phi$, represents the mixing parameter in the BYM2 distribution (see Equation~\eqref{eq:spatialME}). The estimated value highlights that the spatially structured error can actually explain almost 90\% of the total variability observed in the ME model (the credible interval at the 95\% level is equal to [0.895; 0.927]).

\subsection{Car crashes rates}

\begin{figure}
    \centering
    \includegraphics[width=0.8\linewidth]{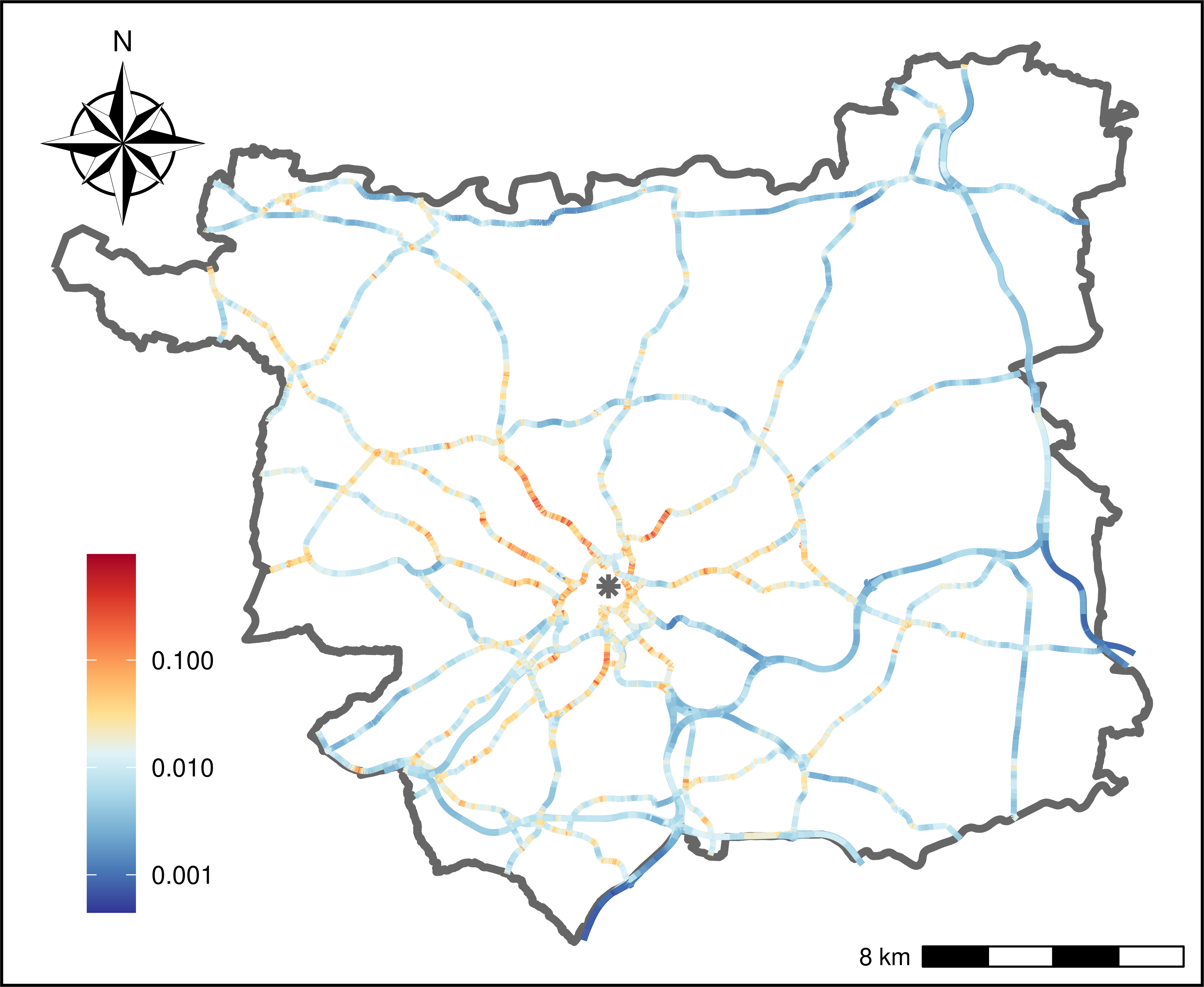}
    \caption{Choropleth map displaying the posterior means of $\lambda_i$ for all street segments in the city network. The colours, which are represented in a logarithmic scale, range from blue (lower risk) to red (higher risk). The grey star denotes the city centre.}
    \label{fig:posterior-rates}
\end{figure}

The results summarised in the previous subsections suggest that the two extensions outperform the baseline model and we compared their fit using the DIC and WAIC criteria introduced in Section~\ref{sec:bayesian-details}. We found that including a spatial random component in the ME structure greatly improves the model's fit (DIC = 64131 vs -98448 and WAIC = 75785 vs -85218); therefore, in the rest of the paper, we will focus on \textit{Extension II}. 

We display in Figure~\ref{fig:posterior-rates} a choropleth map depicting the posterior means of  $\lambda_i$, i.e. the causality rates, for all street segments in the road network. The map was created using a palette of colours going from blue (lower risk) to red (higher risk). Furthermore, considering the skew nature of the posterior rates, the legend is reported in the log scale. The lowest value corresponds to, approximately, one car crash per 9 years every kilometre, while the highest value (which is immensely rare for the data at hand) corresponds to, approximately, 100 car crashes per 9 years every kilometre. Looking at the map, we can clearly recognise the shape of the motorways going through the city (see, e.g., Figure~\ref{fig:tomtom_crashes_counts}), since the corresponding street segments are coloured dark blue. A similar behaviour can be observed in the northern and northeastern parts of the municipality. On the other hand, we can notice that a few segments close to the city centre (identified by a grey star) and several arterial thoroughfares (e.g. \textit{Scott Hall Road}, \textit{Woodhouse Lane}, \textit{Kristall Road}) represent the streets more prone to accidents in the municipality.

\subsection{Model validation}

\begin{figure}
    \centering
    \includegraphics[width=\linewidth]{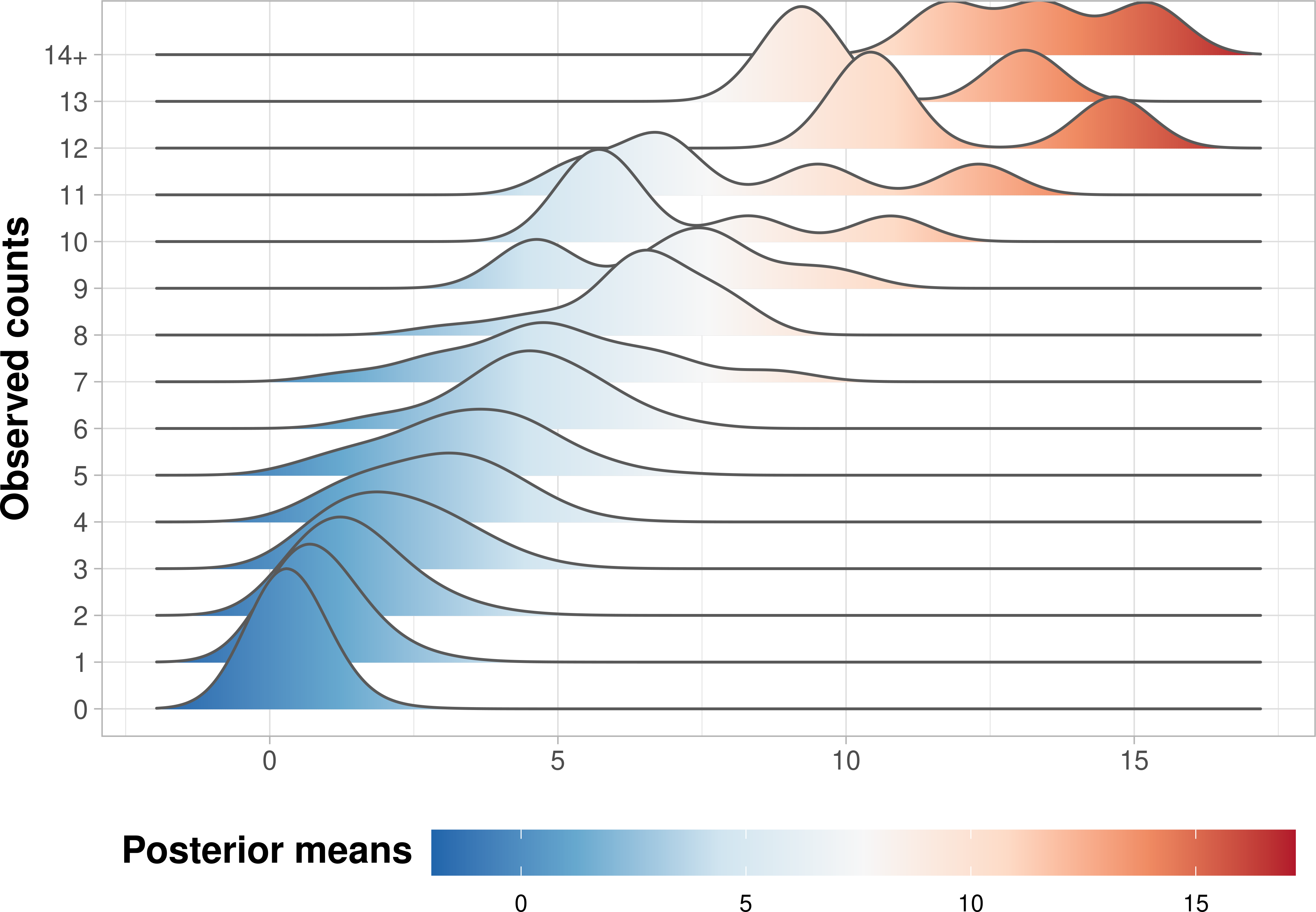}
    \caption{Density curves comparing posterior means of predicted car crashes counts and observed counts. The class \texttt{14+} summarises all street segments that registered fourteen or more car crashes during the years 2011-2019. The values in the last class were lumped since they are extremely sparse in the data at hand.}
    \label{fig:posterior-means}
\end{figure}

The predictive capabilities of the chosen model were validated by computing the posterior means of the car crash rates per segment and comparing these values (multiplied by the corresponding offsets) with the observed counts via a series of density curves. The result is shown in Figure~\ref{fig:posterior-means} and clearly displays a good agreement between the two quantities, being lower car crashes frequencies associated with lower posterior means.  

Finally, considering that the variables included in the regression models are recorded over mismatching time periods (e.g. the network and traffic volumes were collected during the year 2019, the response variable counts the car crashes that occurred from 2011 to 2019, and the socio-demographic covariates were obtained from the 2011 UK census), we tested the temporal stability of our results under the three following scenarios: cases A) and B) consider only the car crashes that occurred during the last three and five years under analysis, respectively; case C) considers the socioeconomic variables pertaining to the 2021 UK Census. In all cases, the other aspects of the model specification remained unchanged. The results are reported in Tables~\ref{tab:fixed-effects-validation} (fixed effects) and~\ref{tab:random-effects-validation} (random effects) where we also compare these estimates with those obtained by \textit{Extension II}. 

The numerical summaries reported in Table~\ref{tab:fixed-effects-validation} show that, in general, the posterior distributions of the fixed effects are stable under different assumptions regarding the temporal aggregation of the variables. In particular, we notice that the magnitude and the significance of the regression coefficients are consistent with \textit{Extension II}. The only situation that slightly differs from \textit{Extension II} is the estimate of the regression coefficient associated to the road traffic variable in scenario A. This is not so surprising since, as already mentioned, it is typically difficult to assess the relationship between road traffic and car crashes risk, particularly when the response variable is extremely sparse as in this case. In fact, during the last three years of the considered period, more than 85\% of street segments in Leeds recorded no car crash. It should also be noticed that, in almost all cases, we obtained a 95\% CI of the regression coefficients (not reported here) that overlaps with the estimates in \textit{Extension II}. 

Table~\ref{tab:random-effects-validation} highlights that scenario C) is virtually equivalent to \textit{Extension II} with respect to the posterior distributions of the hyperparameters. In the other two cases, we observe a slightly higher variability and, in particular, it can be noticed that the posterior means of $\tau_\theta$ exhibits higher values and stronger variation than \textit{Extension II}. This effect can be explained by considering the sparsity of car crashes counts analysed at the network lattice level when the data are aggregated over short temporal periods. 

\begin{table}
    \centering
    \caption{Posterior means and standard deviations (in brackets) of all fixed effects included in \textit{Extension II} and the alternative scenarios described in Section 4.4. The empty cells in column C) pertain to socio-economic variables that, to the best of our knowledge, are not available from the 2021 UK Census data at the time of writing.}
    \label{tab:fixed-effects-validation}
    \begin{tabular}{
    l
    @{}
    S[table-format = +2.3, parse-numbers = false]
    S[table-format = +2.3, parse-numbers = false]
    S[table-format = +2.3, parse-numbers = false]
    S[table-format = +2.3, parse-numbers = false]
    }
    \toprule 
    & {\textit{Extension II}} & {\textit{Case A}} & {\textit{Case B}} & {\textit{Case C}} \\
    \midrule 
    & -11.308 & -11.195 & -15.102 & -13.893 \\
    \multirow{-2}{*}{Intercept} & (0.631) & (1.124) & (1.141) & (0.240) \\
    & 0.743 & 0.616 & 0.921 & 0.915 \\
    \multirow{-2}{*}{FRC 2 (Major roads)} & (0.103) & (0.139) & (0.130) & (0.102) \\ 
    & 0.923 & 0.624 & 1.195 & 1.182 \\ 
    \multirow{-2}{*}{FRC 3 (Secondary roads)} & (0.118) & (0.172) & (0.162) & (0.114) \\
    & 0.045 & -0.017 & 0.025 & -0.060 \\
    \multirow{-2}{*}{Population Density} & (0.064) & (0.065) & (0.064) & (0.052) \\
    & 0.017 & 0.064 & 0.026 & \\
    \multirow{-2}{*}{Young residents (\%)} & (0.131) & (0.133) & (0.132) & \\
    & 0.128 & -0.004 & 0.049 & -0.026\\
    \multirow{-2}{*}{Work from home (\%)} & (0.126) & (0.121) & (0.124) & (0.101) \\
    & -0.059 & 0.355 & 0.102 & 0.055 \\
    \multirow{-2}{*}{Working pop. (\%)} & (0.265) & (0.271) & (0.268) & (0.239) \\
    & 0.159 & 0.097 & 0.084 \\
    \multirow{-2}{*}{Households w/ 4+ vehicles} & (0.132) & (0.131) & (0.131) \\
    & 0.311 & 0.284 & 0.312 & 0.312 \\
    \multirow{-2}{*}{Presence of pedestrian crossings} & (0.047) & (0.071) & (0.057) & (0.048) \\
    & 0.402 & 0.483 & 0.506 & 0.404 \\
    \multirow{-2}{*}{Presence of traffic signals} & (0.050) & (0.075) & (0.060) & (0.051) \\
    & 0.506 & 0.300 & 0.645 & 0.642 \\
    \multirow{-2}{*}{Road traffic} & (0.058) & (0.072) & (0.091) & (0.054) \\
    \bottomrule
    \end{tabular}
\end{table}

\begin{table}
    \centering
     \caption{Posterior means and standard deviations (in brackets) of all hyperparameters included in \textit{Extension II} and the alternative scenarios described in Section 4.4.}
     \label{tab:random-effects-validation}
    \begin{tabular}{
    l
    @{}
    S[table-format = +2.3, parse-numbers = false]
    S[table-format = +2.3, parse-numbers = false]
    S[table-format = +2.3, parse-numbers = false]
    S[table-format = +2.3, parse-numbers = false]
    }
    \toprule 
    & {\textit{Extension II}} & {\textit{Case A}} & {\textit{Case B}} & {\textit{Case C}} \\
    \midrule & 0.066 & 0.328 & 0.144 & 0.062 \\
    \multirow{-2}{*}{$\tau_\theta$} & (0.004) & (0.072) & (0.011) & (0.003) \\
    & 4.471 & 3.780 & 4.457 & 4.693 \\ 
    \multirow{-2}{*}{$\tau_x$} & (0.133) & (0.092) & (0.114) & (0.155) \\
    & 0.120 & 0.061 & 0.077 & 0.143 \\
    \multirow{-2}{*}{$\rho_0$} & (0.024) & (0.012) & (0.019) & (0.020) \\
    & 1.716 & 1.686 & 1.743 & 1.846 \\ 
    \multirow{-2}{*}{$\tau_u$} & (0.105) & (0.130) & (0.114) & (0.114) \\
    & 0.905 & 0.975 & 0.921 & 0.892 \\ 
    \multirow{-2}{*}{$\phi$} & (0.005) & (0.003) & (0.006) & (0.007) \\  
    \bottomrule
    \end{tabular}
\end{table}

\section{Discussion and Conclusions}
\label{sec:conclusion}

In this paper, we considered the problem of adjusting the measurement error in the covariates of a spatial regression estimated at the network lattice level. In particular, we focused on a road safety context and developed a ME model for assessing the relationship between traffic flows and crash frequencies. This is an important (although slightly neglected) topic in the literature. Starting from some relevant papers in the field \parencite{XU2022106518, XIE2018285, KAMEL2020105612, Gilardi2022Multivariate}, we defined a statistical approach that improves over them in the following aspects: 
\begin{itemize}[noitemsep, nolistsep]
    \item the real traffic volumes are approximated using an accurate surrogate derived from GPS devices having global coverage;
    \item the spatial domain is defined using a network lattice, which let us draw conclusions at a granular spatial level that can be extremely informative from a social perspective. Furthermore, analysing accident data at the network level has the additional benefits of being less sensitive to the boundary crash problem and more robust to MAUP \parencite{Gilardi2022Multivariate};
    \item the functional form of the ME model and its key components are specified with great care and informative priors are elicited by means of a set of (geographically sparse but extremely precise) traffic values shared by the DfT. These complementary data are presented in Section~\ref{sec:datacomplementary}. 
\end{itemize}

We tackled the ME analysis problem in a Bayesian framework comparing three increasingly complex hierarchical models. The first one completely ignores the problem of ME, the second one defines a classical ME model at the log-scale level, whereas the last one enhances the correction with a spatially structured random effect. We exemplified the suggested methodologies by analysing the distribution of all car crashes that occurred in the streets of Leeds from 2011 to 2019. We found that, according to DIC and WAIC criteria, the spatial ME (previously denoted as \textit{Extension II}) greatly improves over the classical framework. 

Our results, which are summarised in Section~\ref{sec:results}, highlight the importance of the ME term. In fact, Table~\ref{tab:fixed-effects-log-lambda} shows the posterior means and standard deviations of all fixed effects included in the regression model. Looking at the last row of the table, we can notice the relevance of the attenuation bias, which is typical in presence of measurement errors. Road-specific covariates, namely the road class and the presence of pedestrian crossings or traffic lights, were found to be extremely important. From Table~\ref{tab:fixed-effects-mu} we can deduce that, unsurprisingly, motorways register higher traffic volumes than other road types. Table~\ref{tab:fixed-effects-log-lambda} also highlights two important findings: a) there is a positive correlation between crashes counts and traffic volumes; b) the motorways are the safest road category. These are typical results in the road safety literature. However, we found that the impacts can be seriously attenuated if one estimates the spatial regression model ignoring a proper correction for road traffic volumes. We believe that our results are particularly important from a social perspective since they demonstrate that naive models may provide misleading guidance for policy evaluation if practitioners do not properly take into account these sources of errors. 

Table~\ref{tab:fixed-effects-log-lambda} also suggests that the socio-demographic variables are not related to the car crash occurrences. This is an unexpected finding, which may be due to the overlay operations used to merge the census data with the street segments. In fact, the overlay creates a crude step-wise constant approximation of the socio-demographic data at the network lattice level (i.e. the same value is assigned to all segments in a given LSOA), implying that there might be abrupt changes among neighbouring segments. There exists a vast literature on the spatial misalignment problem for polygonal areas (see, e.g., \textcite[Chapter~7]{banerjee2015hierarchical}), but, to the best of our knowledge, no one extended these methods to the network framework. Model-based approaches may provide a smoother approximation of the census covariates and a better understanding of their relationship with crashes counts. This aspect, despite deserving further investigation, is beyond the scope of this paper since we only focus on the estimation of the relationship between car crashes and traffic counts. Moreover, we also acknowledge that some (or all) of the socio-demographic covariates included into the regression model may suffer from measurement error as well, which would require to be appropriately treated. However, as already mentioned, the main goal of our manuscript is to evaluate the impact of traffic flows on accident frequencies at the network level considering measurements from mobile devices. The proper estimation of a multivariate spatial ME model represents an extremely challenging problem, which is beyond the scope of the present paper and left for future research.

Finally, as reported in Section~\ref{sec:results}, we explored the temporal stability of our results under different time configurations as the one adopted for defining the response variable and the covariates in \textit{Extension II}. In particular, we tested three alternative scenarios that can be characterised as follows: cases A) and B) consider a response variable that aggregates at the street segment level the car crashes that occurred over the last three and five years, respectively; case C) replaces the socio-economic variables with the same covariates pertaining to the 2021 UK Census. Tables~\ref{tab:fixed-effects-validation} and~\ref{tab:random-effects-validation} clearly show that, in almost all situations, the posterior estimates of all elements included in the hierarchical model remain stable under different temporal aggregations.

The approaches presented in this manuscript can be extended in several directions. First, the temporal evolution of car crashes and traffic volumes could be considered, defining a spatio-temporal variation of the suggested models \parencite{boulieri2017space, ma2017multivariate}. Moreover, we could also examine the multivariate nature of the road casualties, computing crashes counts after dividing the events according to one or more characteristics (e.g. the severity level or the number of cars involved) \parencite{barua2014full, Gilardi2022Multivariate}. In the current paper, we decided not to consider any of the two extensions since they introduce a high degree of sparsity in the counts, making the statistical inference even more challenging (as already noticed when analysing scenario A in Section~\ref{sec:results}). These difficulties are even more pronounced when the ME model is coupled with a spatial regression since that correction introduces new parameters, adding even more complexities to the estimation process. 

\section*{Acknowledgements}

We greatly acknowledge the DEMS Data Science Lab for supporting this work by providing computational resources.

\newrefcontext[sorting=nty]
\printbibliography

\end{document}